\documentclass[journal]{IEEEtran}
\ifCLASSINFOpdf
  \usepackage[pdftex]{graphicx}
\else
\fi

\usepackage{xcolor, soul, float}
\sethlcolor{yellow}

\begin{document}
%
\title{HeapSafe: Securing Unprotected Heaps in RISC-V}
%
%
%

\author{Asmit De
        and~Swaroop Ghosh,~\IEEEmembership{Senior~Member,~IEEE}
\thanks{A. De and S. Ghosh are with the School of Electrical Engineering and Computer Science, The Pennsylvania State University, University Park,
PA, 16828 USA e-mail: asmit@psu.edu, szg212@psu.edu}
\thanks{Manuscript received March 18, 2021; revised March 18, 2021.}}

\maketitle

\begin{abstract}
RISC-V is a promising open-source architecture primarily targeted for embedded systems.
Programs compiled using the RISC-V toolchain can run bare-metal on the system, and, as such,
can be vulnerable to several memory corruption vulnerabilities. In this work, we present HeapSafe, a
lightweight hardware assisted heap-buffer protection scheme to mitigate heap overflow and use-after-free vulnerabilities in a RISC-V SoC.
The proposed scheme tags pointers associated with heap buffers with metadata indices and enforces tag propagation for commonly used pointer operations.
The HeapSafe hardware is decoupled from the core and is designed as a configurable coprocessor and is responsible for validating the heap buffer accesses. Benchmark results show a 1.5X performance overhead and 1.59\% area overhead, while being 22\% faster than a software protection. We further implemented a HeapSafe-nb, an asynchronous validation design, which improves performance by 27\% over the synchronous HeapSafe.
\end{abstract}


%
\IEEEpeerreviewmaketitle

\section{Introduction} \label{introduction}
%
%
%
%
\IEEEPARstart{P}{rogramming} languages such as C, which are closer to the hardware, allow direct access to memory and IO to facilitate system and device level programming. Such languages are weakly typed and While being flexible and powerful, this also leads to a plethora of vulnerabilities, if not used with proper practices. C allows memory access using pointers, which are essentially memory addresses that can be referenced and de-referenced for data access. Pointers are an extremely valuable construct as they allow programmers to dynamically allocate memory regions on-demand based on the program's requirement. This saves space and also allows efficient allocation of resources in memory. However, this can also lead to several memory vulnerabilities. Unfortunately, even with decades of research, memory corruption vulnerabilities are still prevalent in modern systems \cite{szekeres2013sok, van2012memory}.

A commonly exploited memory corruption vulnerability is buffer overflow \cite{deckard2005buffer} which occurs when a data is written to a buffer if the size of the data is more than the size of the available buffer. Buffer overflows can occur in a process's stack, or in the heap. A buffer overflow in a process's address space can lead to several exploits such as Control-Flow Integrity violations, Data-Flow Integrity violations, Return-oriented-Programming attacks \cite{abadi2009control, wu2019kepler,castro2006securing, roemer2012return}, etc. Although stack-based buffer overflows are more common, heap buffer overflows can also occur \cite{daniel2008engineering, huang2019analysis, hawkes2008attacking}, and it is more difficult to protect against.

\begin{table}[b]
\caption{Qualitative comparison of Heap Protection Approaches}
\begin{center}
\begin{tabular}{|l|c|c|c|c|}
\hline
\textbf{Type}       & \textbf{Compatibility} & \textbf{Completeness} & \textbf{Performance} & \textbf{Area} \\
\hline
Fat-pointer   & ↓ & ↑ & ↓ & ↓\\
\hline
Low-fat  & ↑ & ↓ & ↑  & ↓ \\
pointer   &  &  &   &  \\
\hline
Object  & ↑ &  ↓  & ↓  & ↓\\
tagging   & &    &   & \\
\hline
Hardware & ↑  &  ↓ & ↑ & ↑ \\
Allocators   &   &   & &  \\
\hline
\textbf{HeapSafe}   & ↑  &  ↑  & ↑ & ↑ \\
\hline
\end{tabular}
\label{comparison}
\end{center}
\end{table}

 

\begin{figure*}[t]
\begin{center}
\includegraphics[trim={0 2.4in 0 0},clip,width=0.7\textwidth]{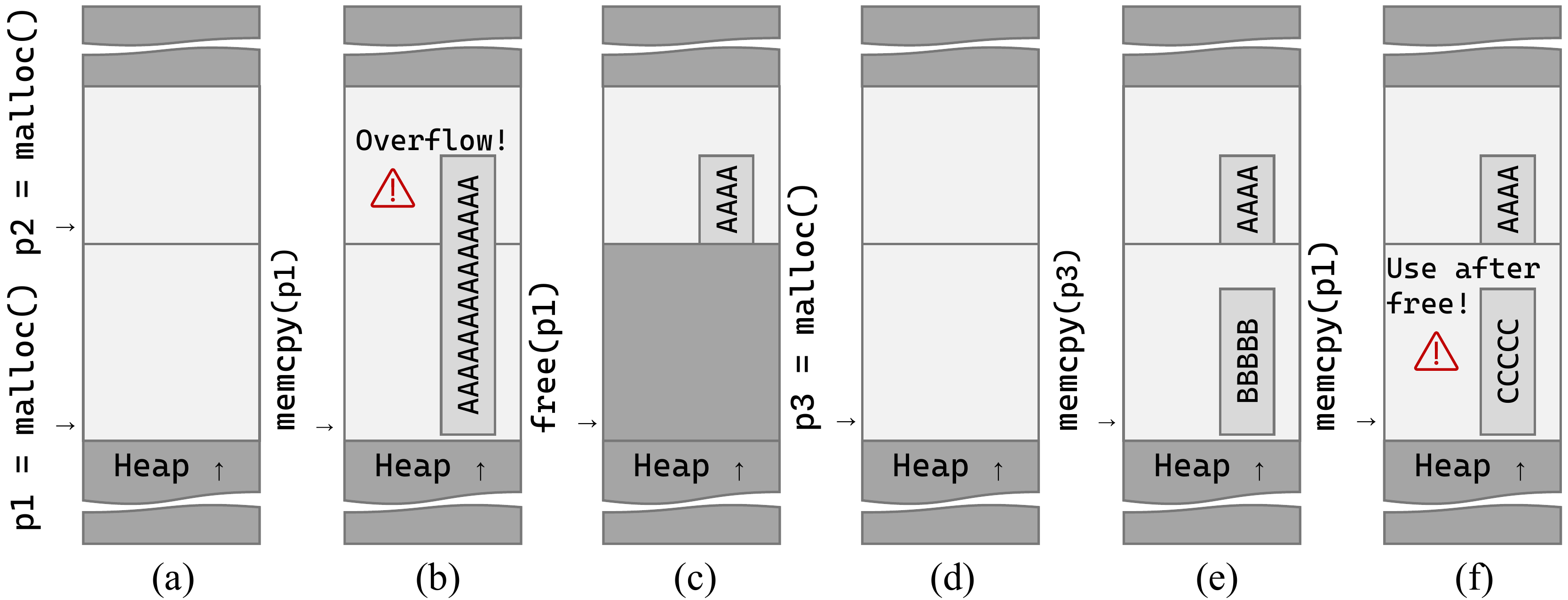}
\end{center}
\caption{(a) Dynamic memory allocation on a heap; (b) Buffer copy and resulting overflow; (c) Memory de-allocation; (d) New allocation in freed location; (e) New data copy to buffer; (f) Data corruption using dangling pointer (use-after-free).}
\label{heap}
\end{figure*}

A heap buffer overflow occurs when a pointer used to write data to the buffer goes beyond the allocated region of the heap and overwrites the critical data, potentially allowing an adversary to launch attacks, such as, write-what-where, malicious shellcode execution, etc. This is possible due to the lack of bounds checking (a technique that allows validation of a pointer's access bounds) in pointers. Unlike a stack-buffer overflow based attack, which occurs in tandem with the rolling/unrolling of the process's stack, heap-buffer overflows are arbitrary and can happen anywhere at any point during the program's execution. Stack-buffer based overflows, and consequently ROP attacks can be mitigated using techniques such as stack canaries \cite{cowan1998stackguard}, shadow stacks \cite{burow2019sok}, etc. However such techniques are not applicable to heap-based buffer overflows, since the allocated memory is dynamic and does not follow the process's stack frame, and as such, there are no return addresses to protect. This makes heap-buffer overflows much harder to detect and prevent.

Several techniques have been explored in literature to mitigate heap buffer overflows. Pointer protection is  a common technique that provides protection to pointers based on bounds checking of allocated memory \cite{necula2005ccured, grossman2005cyclone}. Such techniques associate additional metadata structures with the pointers and provide software runtimes to perform access validation. A similar metadata association approach has also been explored for objects instead of pointers \cite{akritidis2009baggy}. A better alternate approach to pointer-based protection is low-fat pointers \cite{duck2016heap}, where instead of using additional metadata structures with pointers, the authors have utilized the native pointer itself for storing the metadata, thereby preserving backwards compatibility. A few hardware based memory allocation and runtime monitoring approaches have also been explored in \cite{arias2017ha2lloc, arora2005secure}.  Recently, there has been some developments on a secure and memory safe processor \cite{shakti-t, shakti-ms}, which utilizes the fat-pointer scheme implemented in the architecture pipeline. The primary difference between Shakti-T \cite{shakti-t} and our work is that, Shakti-T is a processor with memory safety built into the pipeline, and hence is not scalable or customizable. The metadata field width is also fixed. Our implementation is a decoupled implementation which can be attached to any RISC-V core interfacing with the RoCCIO. Since our implementation is on a custom coprocessor, the design can be tuned and scaled according to needs, without modification of the core pipeline architecture. Table \ref{comparison} provides a qualitative analysis of the different heap protection approaches.

RISC-V is a promising open source instruction set architecture (ISA) that can be adapted to SoC architectures targeting a varied range of applications such as, IoT devices, machine learning accelerators and even data-center microprocessors. It allows programs and applications to run bare-metal on the hardware for application specific scenarios. Such applications has complete access to the range of memory available in the hardware, and as such, can suffer from the same memory corruption vulnerabilities.

We propose HeapSafe, a hardware assisted heap protection engine built on the RocketChip SoC \cite{lee2015risc} running the RISC-V ISA. We leverage the Rocket Custom Coprocessor (RoCC) to design the HeapSafe module for protection from heap-buffer overflows and use-after-free attacks. Over a traditional software based approach, HeapSafe incurs no performance losses due to context switching or cache replacement, no manual secure memory management for bare-metal applications and no compiler dependent pointer analysis. HeapSafe is able to achieve high backwards compatibility and completeness, while retaining good performance.

\section{Background} \label{background}

In this section, we explain two types of memory corruption attacks on the heap, against witch HeapSafe can be applied. We also describe the RISC-V RocketChip SoC platform and the Rocket Custom Coprocessor, which is our target implementation platform.

\subsection{Heap attacks}
A heap buffer is a memory space dynamically allocated on a process's heap. In the userspace, a heap is created using GNU C library (glibc) functions such as, \texttt{malloc()} or \texttt{calloc()} as shown in Fig. \ref{heap}. The function returns the memory address of the first byte of the allocated space, and is used as the pointer to the heap. Data is written to the heap's allocated bytes with the pointer using glibc functions such as \texttt{memset()}, \texttt{memcpy()}, \texttt{strcpy()}. In the following paragraphs, we explain the basics of heap-based attacks using known vulnerabilities from the CWE database.

\begin{figure}[b]
\begin{center}
\includegraphics[trim=0 2.9in 0 0,clip,width=0.49\textwidth]{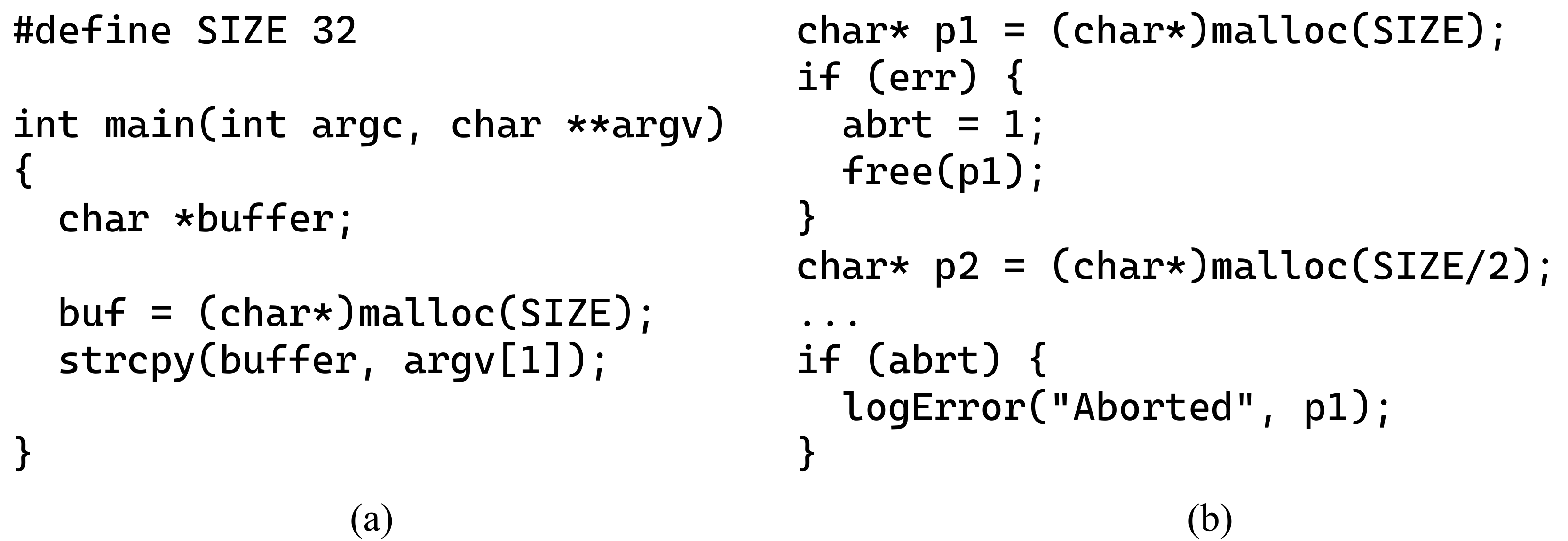}
\end{center}
\caption{Vulnerable code showing (a) heap buffer overflow weakness; (b) use-after-free weakness.}
\label{attack}
\end{figure}

\textbf{Buffer overflow (CWE-122)}: Fig. \ref{attack}(a) shows a simple code demonstrating a buffer overflow vulnerability on the heap. Here, a heap buffer is allocated on the process's memory and a string from the command-line is copied to the buffer. However, it is easy to overflow the buffer if the size of the string is more than \texttt{SIZE}. This can potentially overwrite process data on the heap in other allocated buffers and may lead to memory exploits. The scope of such attacks is quite large, since, in real applications, a lot of complex constructs such as objects, structs, function pointers, etc. are allocated on the heap.

\textbf{Use after free (CWE-416)}: This vulnerability occurs when, heap memory is reallocated after the data on a heap is freed. A previously leftover reference (dangling pointer) to that memory can potentially access newly created data from that heap location. Fig. \ref{attack}(b) shows an example of this vulnerability. In this case, location pointed by \texttt{p1} gets freed if an error occurs. It is possible that when memory referred by \texttt{p2} is created, it is at the same freed location as \texttt{p1}. In such a situation, referring to \texttt{p1} later in the code can inadvertently leak or even corrupt data at that location.

\subsection{RocketChip System and RoCC}

\begin{figure}
\begin{center}
\includegraphics[trim={1.2in 0 1.8in 0},clip,width=0.49\textwidth]{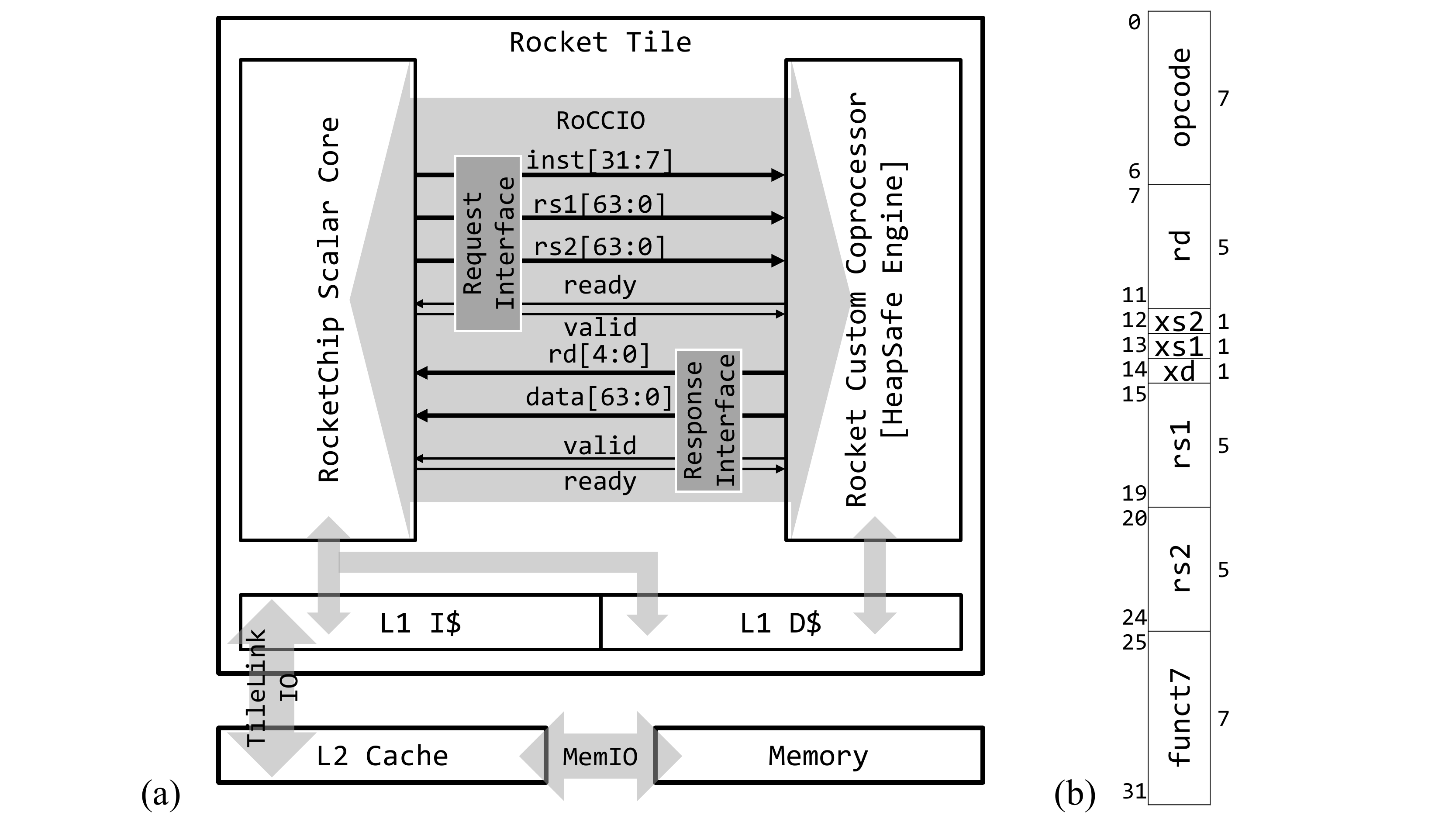}
\end{center}
\caption{(a) RocketChip SoC architecture with HeapSafe implemented RoCC; (b) RoCC instruction encoding.}
\label{rocketchip}
\end{figure}

The HeapSafe architecture is based on Rocket Chip (written in CHISEL), an open source parameterized system-on-chip (SoC) design generator. We use the RocketChip generator to generate synthesizable RTL for the standard Rocket Core SoC, a six-stage single-issue in-order pipeline processor that executes the 64-bit scalar RISC-V ISA (Fig. \ref{rocketchip}(a)). The Rocket Tile consists of the scalar core, the L1 caches, and the Rocket Custom Coprocessor (RoCC). The RoCC is a user-defined accelerator for the core which communicates with core over the RoCCIO interface using a set of custom instructions.

\textbf{RoCC Instructions:} The 32-bit RoCC instructions extend the RISC-V ISA and are encoded as shown in Fig. \ref{rocketchip}(b). The four custom instructions supported by Rocket Chip are \textit{custom0-3}, each having a different opcode. The \textit{xs1}, \textit{xs2}, and \textit{xd} bits control read/write of the core registers by the RoCC instruction. If \textit{xs1} is 1, the 64-bit value in the register specified by \textit{rs1} is passed to the RoCC. Similarly, \textit{xs2} bit controls the read of register specified by \textit{rs2}. If \textit{xd} bit is 1 and \textit{rd} is not 0, the core will wait for a value to be returned by the coprocessor over RoCCIO after issuing the instruction to the coprocessor. The value is then written to the register specified by \textit{rd}. If the \textit{xd} is 0 or \textit{rd} is 0, the core will not wait for a value from RoCC. The opcode field specifies the custom instruction for the RoCC, and the \textit{funct7} field further specifies a user-defined function implemented in the RoCC. The RoCC is responsible for signaling illegal instructions to the core. 

\textbf{RoCCIO Interface:} The RoCC interacts with the Rocket core and the shared memory system via the RoCCIO interface. The core initiates a RoCC command by passing the RoCC instruction to the coprocessor via inst, as well as the relevant register values via \textit{rs1} and \textit{rs2}. If the RoCC instruction has the \textit{xd} bit set, then the RoCC must eventually supply a response value over the RoCC response interface via data.

\section{HeapSafe Implementation for RISC-V} \label{implementation}
In this section we describe the implementation of the HeapSafe engine, the associated HeapSafe library and scope of protection. We also explain the usage of HeapSafe in standard C programs for heap buffer protection.

\subsection{HeapSafe protection scope}
In this work, we aim to protect dynamically allocated buffers on the heap that are accessed using pointers derived from the allocation pointer. We enforce metadata propagation between pointers, and the system is able to trace the correct metadata for all pointer arithmetic operations. HeapSafe is able to protect against buffer overflow on heap, and also prevent inadvertent use-after-free accesses. Any program targeted for the RISC-V system can be updated to use the safe heap functions from the HeapSafe library. Each pointer used to allocate a heap buffer will be converted to a \textit{safe\_pointer} by the HeapSafe library. Any other pointers derived from the \textit{safe\_pointer} is also tagged as a \textit{safe\_pointer}. The tag is an identifier for the pointer that is encoded in the higher order bits of the pointer. We enforce tag propagation between pointers for all pointer assignments and pointer arithmetic operations. This allows us to propagate pointer metadata information across pointers. The program is compiled by including the HeapSafe library and while running on the core, the HeapSafe hardware is responsible for storing and validating heap metadata.

Since we are reusing the same pointer to store the tag, referencing pointers is trivial, without having to process the pointer information. This also allows us to easily enforce tag propagation in the following scenarios (Table \ref{tagprop}):

\begin{table}
\caption{HeapSafe Tag Propagation}
\begin{center}
\begin{tabular}{|c|l|}
\hline
\textbf{Case}       & \textbf{Code} \\
\hline
Memory allocation   & \texttt{safe\_ptr = safe\_malloc(size);} \\
\hline
Assignment          & \texttt{safe\_ptr2 = safe\_ptr1;} \\
\hline
Pointer             & \texttt{safe\_ptr2 = safe\_ptr1 + offset;} \\ 
Arithmetic          & \texttt{safe\_ptr2 = safe\_ptr1 - offset;} \\
\hline
Type cast           & \texttt{safe\_ptr2 = (type*) safe\_ptr1;} \\
\hline
\end{tabular}
\label{tagprop}
\end{center}
\end{table}

(a) \textit{Memory allocation:} When allocating a heap buffer, a new \textit{tag} is generated. The \textit{safe\_pointer} is created using the \textit{tag} and the base pointer (\textit{raw\_pointer}).

(b) \textit{Pointer assignment:} During a pointer assignment, the \textit{tag} from the original \textit{safe\_pointer} is propagated to the new \textit{safe\_pointer} alongside the \textit{raw\_pointer} value.

(c) \textit{Pointer arithmetic:} During a pointer arithmetic operation such as array access at a specific index, the new \textit{safe\_pointer} created by adding/subtracting the offset receives the same \textit{tag} as the original \textit{safe\_pointer}.

(d) \textit{Pointer type conversion:} When a \textit{safe\_pointer} is cast to a new type, the \textit{tag} is propagated to the \textit{safe\_pointer} of the new type.

Aside from these four cases, storing and retrieving \textit{safe\_pointers} from memory are trivial and same as storing and retrieving normal pointers from memory. The \textit{tag} in the \textit{safe\_pointer} is retained throughout the store and retrieve operations. Passing \textit{safe\_pointers} as function arguments and returning \textit{safe\_pointers} from functions are also same as with normal pointers, and the\textit{tag} is retained in the process. Two other cases that need special mention are the null pointer and manual pointer creation from integer values. In both cases, the \textit{tag} is set to 0. The \textit{tag} 0 is also indicates that the pointer is not protected and any safe heap operations using the HeapSafe library with the pointer will result in an error.

\subsection{HeapSafe library}
The HeapSafe protection engine is accompanied by a library containing safe implementations of critical heap buffer functions such as \texttt{safe\_malloc()}, \texttt{safe\_copy()}, \texttt{safe\_free()} and \texttt{safe\_read()} / \texttt{safe\_write()}, that utilize the HeapSafe hardware. We describe the operation of these helper functions below.

\subsubsection{\textbf{\texttt{safe\_malloc()}}} 
\begin{figure}
\begin{center}
\includegraphics[trim={2in 6.7in 2in 0},clip,width=0.49\textwidth]{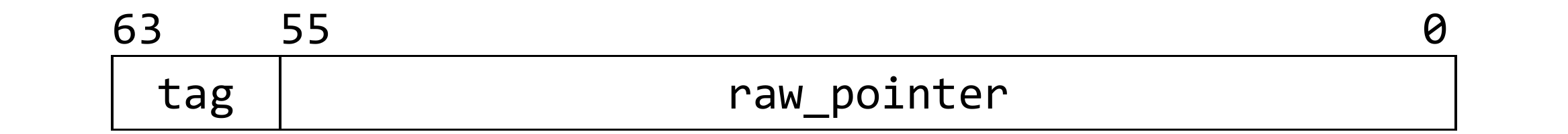}
\end{center}
\caption{\textit{safe\_pointer} bit allocation scheme.}
\label{safeptr}
\end{figure}

This function allocates a buffer in the process's heap similar to \texttt{malloc()} in the GNU C library. The allocated memory is referred to by the address of the first byte of the heap called the \textit{raw\_pointer}. We create a \textit{safe\_pointer} from the \textit{raw\_pointer} by using the top most significant byte as a tag reference. The bit allocation is shown in Fig. \ref{safeptr}. We assign the \textit{tag} from a static list of available tags, which are local to the process. Since we are using 1 byte to represent the \textit{tag}, we can use HeapSafe for a maximum of 255 simultaneous heap allocations in the process. We exclude 0 as a \textit{tag} to maintain compatibility with pointers that are not using the HeapSafe engine. Furthermore, this also excludes the higher order 256 byte memory region in the address space, however, this allocation scheme is sufficient for standard RISC-V applications running bare-metal. For RISC-V systems with user-space and kernel-space separation in memory, HeapSafe is able to protect user-space processes only. It is to be noted that, even though we have used 1 byte for representing the \textit{tag}, this bit allocation is customizable in the library. While compiling the HeapSafe library, the bit allocation for the \textit{tag} can be set as required to match with the HeapSafe hardware (details in Section \ref{conf}).

After allocating the \textit{tag}, we send a custom RoCC instruction \texttt{HS\_STORE} to the HeapSafe hardware to write the pointer metadata. We send the \textit{safe\_pointer} and the size of the allocated buffer encoded in the RoCC instruction.

The \texttt{HS\_STORE} instruction is crafted as follows: The \textit{opcode} is set to \textit{custom0} (b'0001011), \textit{rs1} is set to the register containing the \textit{safe\_pointer}, \textit{rs2} is set to the register containing the size of the heap buffer, \textit{xs1} and \textit{xs2} fields are set to 1, and \textit{funct7} is set to \textit{hs\_store} (b'0000000). The instruction is non-blocking, and the program proceeds without waiting for a response from the HeapSafe engine.

\subsubsection{\textbf{\texttt{safe\_copy()}}} This function enables a safe copy operation from a source buffer to the destination buffer which guarantees that the buffer will not be overflowed. To perform the copy operation, we check the pointer being used to refer to the destination heap buffer. We send the pointer with the \texttt{HS\_VALIDATE} instruction to the HeapSafe engine to perform an out-of-bounds validation.

The \texttt{HS\_VALIDATE} instruction is crafted as follows: The \textit{opcode} is set to \textit{custom0} (b'0001011), \textit{rs1} is set to the register containing the pointer, \textit{rd} is set to a register to receive the validation outcome, \textit{xs1} and \textit{xd} fields are set to 1, and \textit{funct7} is set to \textit{hs\_validate} (b'0000001).

After performing the out-of-bounds validation, the HeapSafe engine returns a 0 or 1 indicating in-bounds or out-of-bounds respectively. The \texttt{safe\_copy()} function can then proceed or halt based on the validation outcome.

\subsubsection{\textbf{\texttt{safe\_free()}}} This function is complementary to \texttt{safe\_malloc()} which allows a clean de-allocation of the memory space and metadata removal from the HeapSafe engine. We first parse the \textit{safe\_pointer} to extract the \textit{raw\_pointer}. We then send the \textit{safe\_pointer} to the HeapSafe engine with \texttt{HS\_FREE} instruction to perform the metadata removal.

The \texttt{HS\_FREE} instruction is crafted as follows: The \textit{opcode} is set to \textit{custom0} (b'0001011), \textit{rs1} is set to the register containing the \textit{safe\_pointer}, \textit{xs1} is set to 1 and the \textit{funct7} field is set to \textit{hs\_free} (b'0000011). The instruction is non-blocking, so the program continues to execute on the core without waiting for a response from the HeapSafe engine. After sending the \texttt{HS\_FREE} instruction, the memory de-allocation is performed normally, similar to \texttt{free()} in glibc.

\subsubsection{\textbf{\texttt{safe\_read()}} / \textbf{\texttt{safe\_write()}}} By re-purposing the MSB bits of the original pointer to store the \textit{tag}, we get the benefit of enforcing easy tag propagation. However, it precludes us from performing pointer de-referencing to read/write data in the standard way. This is because, the \textit{safe\_pointer} by itself is not a valid memory address due to the inclusion of the \textit{tag} bits, and de-referencing in the usual way, e.g., \texttt{data = *safe\_ptr;} will raise a memory access exception. To circumvent this issue, we have also provided \texttt{safe\_read()} and \texttt{safe\_write()} functions, that can safely extract the \textit{raw\_pointer} from the \textit{safe\_pointer}. It then sends a \texttt{HS\_VALIDATE} instruction to HeapSafe engine to validate the read/write access, and then performs the de-referencing to read/write data in memory based on the \textit{raw\_pointer}:

\begin{verbatim}
    addr = extractRawPointer(safe_ptr);
    data = *addr; // For read
    *addr = data; // For write
\end{verbatim}

\subsection{HeapSafe hardware}

\begin{figure*}
\begin{center}
\includegraphics[trim=0 0.4in 0 0,clip,width=0.65\textwidth]{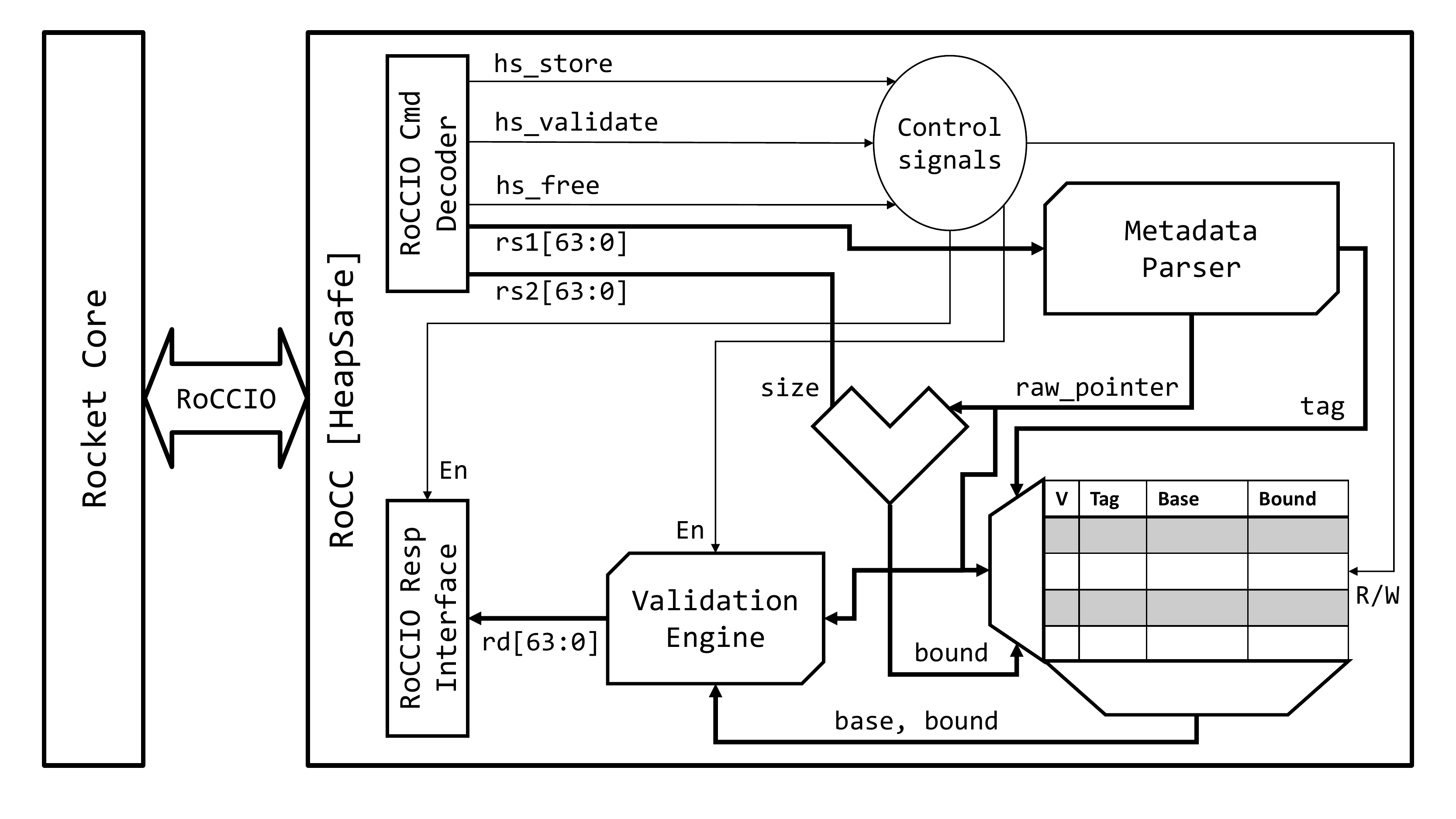}
\end{center}
\caption{HeapSafe architecture in RoCC.}
\label{heapsafe}
\end{figure*}

The HeapSafe engine is a custom designed accelerator that is decoupled from the processor core and connected over the RoCCIO interface. The engine consists of a metadata parser, a metadata table, and a validation engine as shown in Fig. \ref{heapsafe}.

The HeapSafe engine receives commands over the RoCCIO request interface. The Cmd Decoder decodes the RoCC instruction to read the \textit{opcode}, the \textit{rs1} an \textit{rs2} data fields, and the \textit{funct7} function field, and asserts the required control signals. In our implementation the opcode field is always decoded to \textit{custom0}. The \textit{rs1} and \textit{rs2} data fields contains pointer metadata as requires for a specific function. The \textit{funct7} field is decoded to functions such as, \textit{hs\_store}, \textit{hs\_validate} and \textit{hs\_free}.

\textbf{\textit{hs\_store}:} This function indicates a heap buffer creation and instructs the HeapSafe engine to store the associated metadata which is received in the form of the \textit{safe\_pointer} on the rs1 field and the size on the rs2 field. The metadata parser processes the \textit{safe\_pointer} and extracts the tag and the \textit{raw\_pointer} values based on the specified bit encoding.

The metadata table is implemented as a hardware content-addressable memory for parallel search and fast lookup. The table consists of three fields for storing the metadata - (i) Tag, (ii) Base, and (iii) Bound. Each row in the table is designed as a vector of the three 64-bit wide fields. In addition to the metadata, each row in the table also contains a valid bit to indicate the validity of the metadata. The size of the table is customizable as part of the design and can be set while instantiating the hardware. In our implementation for testing, we have used a table consisting of 256 rows, allowing metadata storage for a maximum of 256 heap buffers.

A write signal is automatically issued on decoding the \textit{hs\_store} as the function, which writes the pointer metadata at an available location in the metadata table. These locations are indicated by the valid bit set to 0. The parsed tag is stored in the Tag field and the raw\_pointer is stored in the Base field. Instead of storing the size of the heap buffer, we pre-compute the bound address value as:
\begin{equation}
Bound[Tag] = raw\_pointer + size \label{eq1}
\end{equation}
Since the metadata store instruction is non-blocking, we improve performance by saving the calculated bound address value in the Bound field. Finally, the valid bit is set to 1 to mark the row as active.

\textbf{\textit{hs\_validate}:} This function indicates a heap buffer write operation on the core and instructs the HeafSafe engine to validate the pointer's access bounds. The pointer being used to access the heap is received on the \textit{rs1} field. The metadata parser processes the pointer on the \textit{rs1} field and extracts the current \textit{tag} and the \textit{raw\_pointer} (memory address) values.

A read signal is issued on decoding the \textit{hs\_validate} as the function, which performs a parallel search of the current \textit{tag} on the \textit{Tag} field of the metadata table. Once a tag match is found, the \textit{Base} and \textit{Bound} fields are read. The access bounds for the current pointer (\textit{ptr}) is validated as:
\begin{equation}
isOOB = (ptr < Base) \parallel (ptr \ge Bound) \label{eq2}
\end{equation}
The out-of-bound signal (\textit{isOOB}) is asserted when the current pointer (memory address) is either less than the lower bound (base), or is greater than or equal to the upper bound of the heap buffer. The value of the \textit{isOOB} signal is held at 0 if the current pointer is within bounds. Since the validation is a blocking operation by default, having the upper bound address of the heap buffer in the metadata table speeds up the validation time. The value of \textit{isOOB} signal is placed on the \textit{rd} field of the response interface and is sent back to HeapSafe library to take the required action - (i) proceed or (ii) terminate.

\textbf{\textit{hs\_free}:} This function indicates a heap buffer de-allocation on the core, and instructs the HeapSafe engine to clear its corresponding metadata. However, instead of removing or zero-izing the metadata from the table, we set the valid bit for the row to 0 to invalidate the metadata entry and mark it as available for future use. The \textit{hs\_free} operation is non-blocking and the program on the core continues to run without waiting for a response from the HeapSafe engine.
Invalidating metadata entries allow us to mitigate inadvertent use-after-free vulnerabilities, since the \textit{tag} for the dangling pointer will be invalidated after free. 

\subsection{HeapSafe usage in C programs} \label{usage}
\begin{figure}
\begin{center}
\includegraphics[trim=0 2.1in 0 0,clip,width=0.49\textwidth]{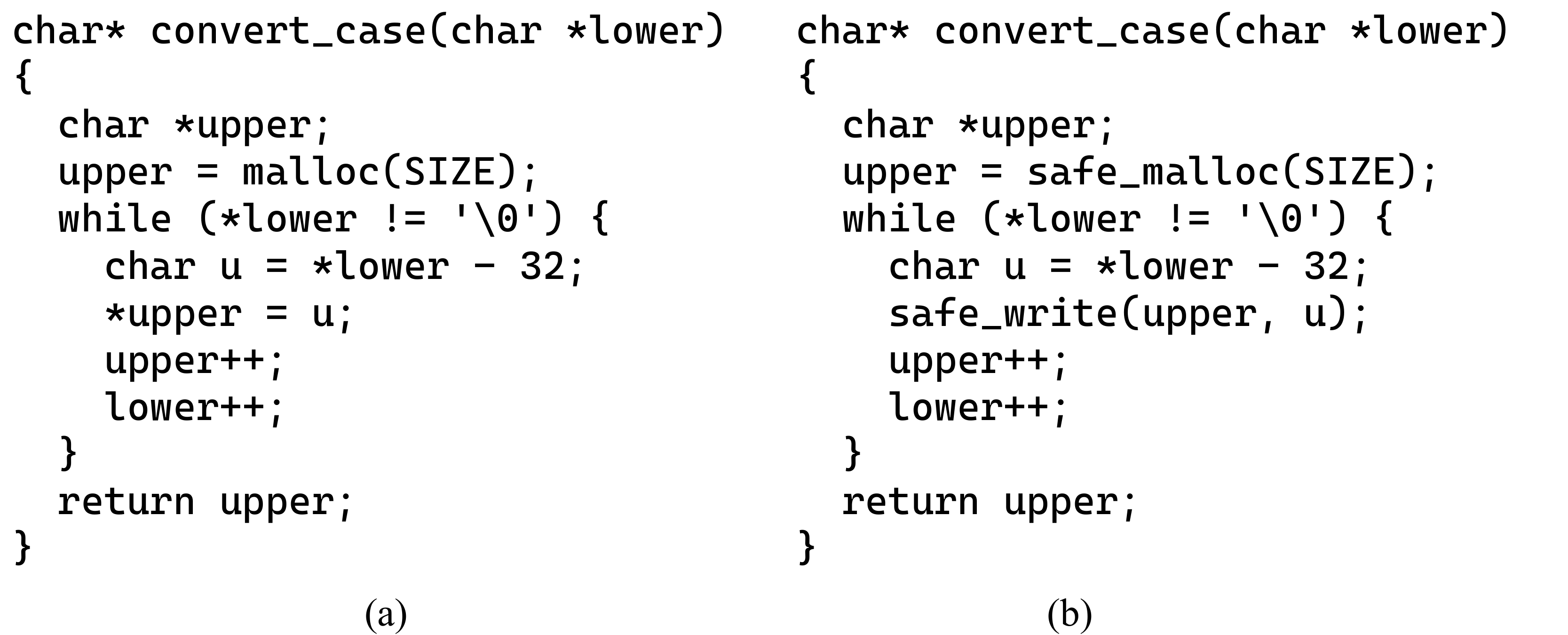}
\end{center}
\caption{(a) Source code with heap buffer overflow vulnerability; (b) HeapSafe protected code to prevent overflow.}
\label{hsuse}
\end{figure}

We demonstrate a basic use of HeapSafe in a simple C program (Fig. \ref{hsuse}a). Let us consider a function that receives a lowercase string data, converts each character to uppercase and stores to a buffer allocated on the heap. The function then returns the pointer to the heap buffer storing the uppercase string.

In this program, the \texttt{upper} buffer is vulnerable to overflow, and hence, let us modify the code to use HeapSafe to protect the buffer. We assume that \texttt{lower} buffer doesn't need to be protected. We perform the modification as follows:

We first replace the \texttt{malloc()} function call with \texttt{safe\_malloc()}, that returns a tagged \texttt{safe\_pointer} to \texttt{upper}. When iterating over the characters from the source buffer, after converting to lowercase, we need to de-reference the \texttt{upper} pointer to store the uppercase character. We modify the de-referencing code with the \texttt{safe\_write()} function that performs the correct write to memory operation. Towards the end of the loop, we update the destination heap pointer (\texttt{upper}) by increasing the pointer by 1. The pointer arithmetic propagates the original tag for \texttt{upper} to the new \texttt{upper}. The HeapSafe protected code is listed in Fig. \ref{hsuse}b.

\section{Evaluation} \label{evaluation}

\begin{figure}
\begin{center}
\includegraphics[trim=0 3.2in 0 0,clip,width=0.49\textwidth]{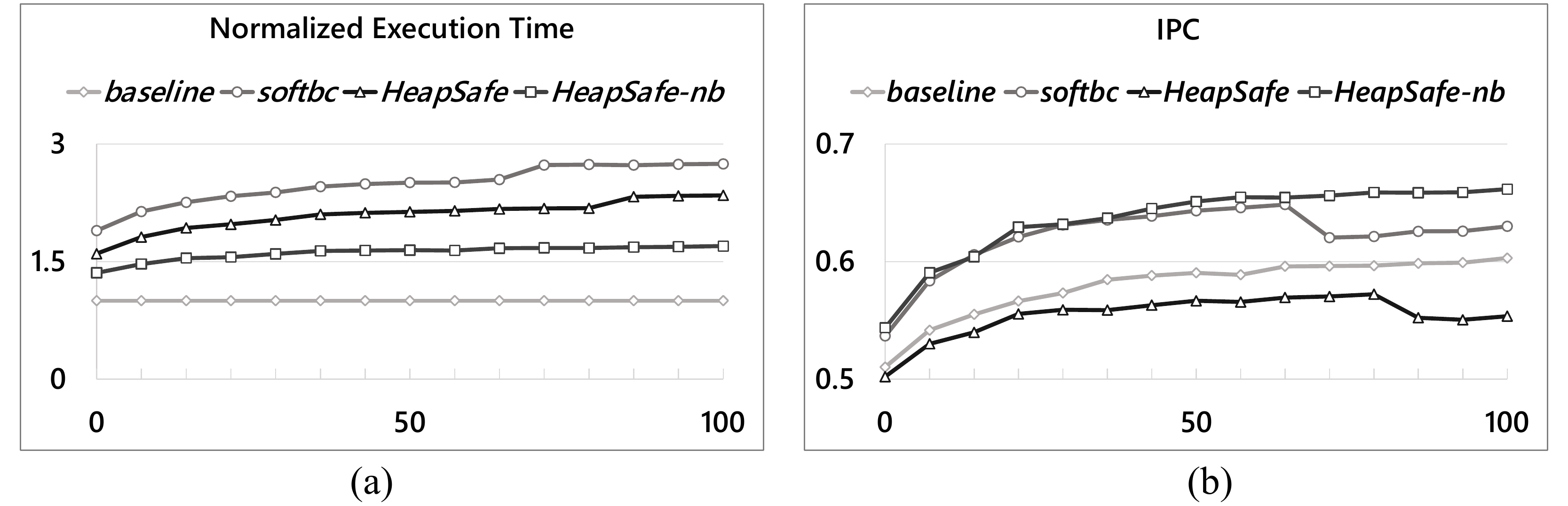}
\end{center}
\caption{(a) Execution time trend normalized to baseline; (b) IPC trend. X-axis represents the percentage of buffer copy operations occurring on the heap.}
\label{simtrend}
\end{figure}

\begin{figure}
\begin{center}
\includegraphics[trim=0 3.2in 0 0,clip,width=0.49\textwidth]{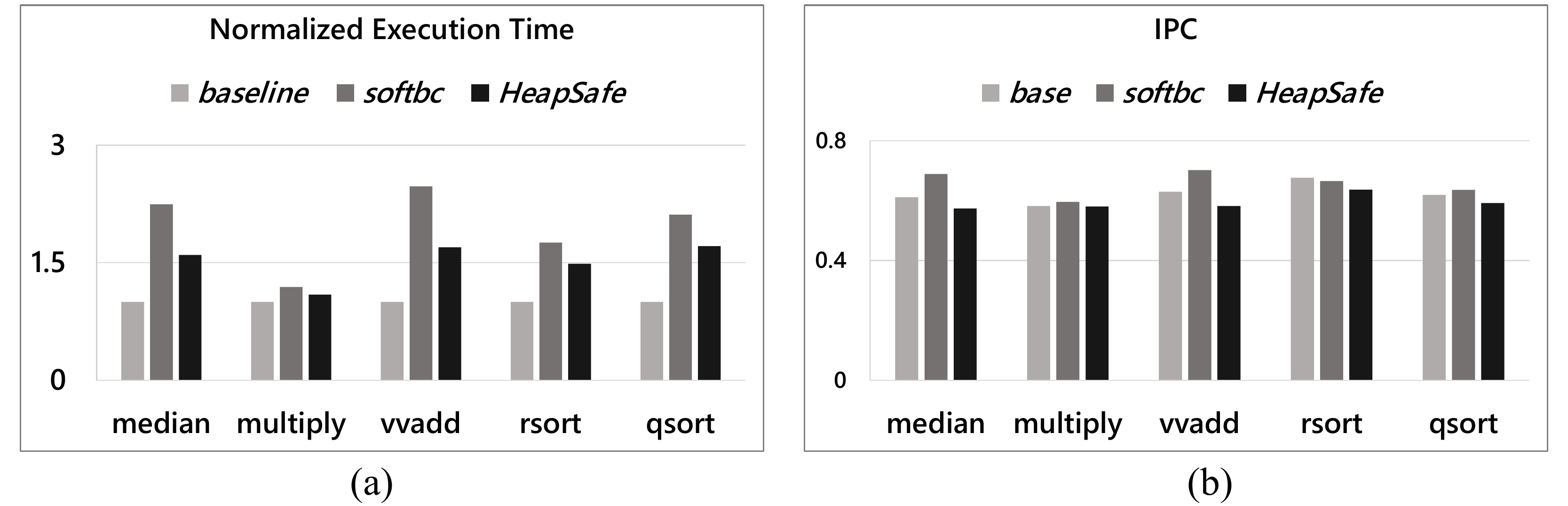}
\end{center}
\caption{(a) Execution time trend normalized to baseline; (b) IPC trend. X-axis represents the percentage of buffer copy operations occurring on the heap.}
\label{bmarks}
\end{figure}

We evaluated HeapSafe by generating a RocketChip SoC design config with the HeapSafe module. We tested the HeapSafe security architecture in the C++ cycle accurate emulator built from the config. The hardware architecture of HeapSafe is coded in CHISEL and synthesizable verilog is generated using the RocketChip generator. To evaluate performance, we created sample workloads that perform multiple buffer copy operations on the process's stack and heap. We compiled three versions of the code: (i) \textit{baseline} with no protection, (ii) \textit{softbc} with in-process software-based bounds checking, and (iii) \textit{HeapSafe} with our HeapSafe library and protections. We swept the workload balance between the stack and the heap and evaluated the trend as shown in Fig. \ref{simtrend}. We note that \textit{HeapSafe}'s execution time overhead is low and similar to \textit{softbc} when there are more stack workloads; however, as the heap workload increases compared to the stack workload, \textit{HeapSafe} tends to perform better. At around 75\% heap workload, \textit{HeapSafe} performs 20\% faster than \textit{softbc}. However, instructions-per-cycle (IPC) suffers in \textit{HeapSafe} compared to \textit{softbc}, since \textit{HeapSafe} runs less instructions in total. We have also implemented a fully non-blocking version of HeapSafe (details in Section \ref{pvs}) which outperforms \textit{softbc} in both execution time and IPC at the cost of delayed heap corruption detection. At high heap workloads, \textit{HeapSafe-nb} is 38\% faster than \textit{softbc}. We estimated the area of HeapSafe by generating a E300 Arty FPGA bitstream and found it to have a nominal 1.59\% overhead (number of cells) over the default configuration. We further updated the RISC-V ISA test benchmarks with \textit{HeapSafe} and \textit{softbc} and evaluated their performance (Fig. \ref{bmarks}). \textit{HeapSafe} incurs a 1.5X overhead over \textit{baseline} on average, while being 22.4\% faster than \textit{softbc}. Average IPC is slightly low at 0.59 compared to 0.62/0.65 (\textit{baseline}/\textit{softbc}). A qualitative evaluation of HeapSafe has been shown in Table \ref{comparison} as well.

\section{Discussions} \label{discussions}

\subsection{HeapSafe design improvements} \label{pvs}
HeapSafe implements fine-grained approach for instant detection of heap overflow at the cost of some performance penalties. However, such high security guarantees are not needed in less critical systems. We propose the following alternate flavors of HeapSafe system design:

\subsubsection{Non-blocking validation}
The validation mechanism with the \texttt{HS\_VALIDATE} instruction can be made non-blocking to avoid wait by the core for a response from HeapSafe. In this design, HeapSafe will asynchronously raise an exception on the core when it detects an out-of-bounds error (instead of sending the value of the \textit{isOOB} signal on the response interface). An exception handler is implemented in the HeapSafe library will terminate the application in such cases. This approach will improve performance by 27.6\% (Fig. \ref{simtrend}(a)) at the cost of slight delay in attack detection.

\subsubsection{Compiler support} 
HeapSafe hardware is currently complemented by the HeapSafe library to replace unsafe heap operations with safe variants. 
Replacement of unsafe heap operations with safe variants using library can be automated by adding compiler support. In this design, we compile the source program to be protected using LLVM/Clang for RISC-V. The LLVM generates an intermediate representation (IR) of the source code. An LLVM compiler pass is written to parse the IR to scan for heap pointers and replace them with the tagged \textit{safe\_pointers}. The custom HeapSafe instructions are also inserted for the required operations to communicate with the HeapSafe engine. The IR is then compiled to generate the ELF binary to run on the system. This improvement is more design friendly to the source code programmer since the code does not need to be explicitly updated to use the HeapSafe engine. 

\subsubsection{Top byte ignore} 
Pointer de-referencing requires additional library functions to perform the correct pointer extraction in HeapSafe. This can be avoided if the core is set to ignore the top byte for any memory address in the user-space. This can be achieved by conditionally masking the top byte of the address in the address decoder in the core pipeline. This will guarantee that any tagged \textit{safe\_pointer} is seen as a normal \textit{raw\_pointer} in the pipeline, and all load/store operations will automatically be performed using the \textit{raw\_pointer}.

\subsubsection{Multi-process support}
Our current HeapSafe implementation is targeted to protect a single process, or a process running bare-metal on the system. 
However, due to the decoupled coprocessor based design, HeapSafe can be scaled up to support protection of multiple processes simultaneously. RISC-V cores view each running process as a hardware thread (hart). In the scaled up implementation, multiple instances of the HeapSafe coprocessor is instantiated in the same tile along with the core. Each instance of HeapSafe engine is associated with a \texttt{hartId}. When a process using the HeapSafe library is running on the core, the system hardware selects the specific HeapSafe engine to use with the hart for that process.

\subsection{Security of HeapSafe hardware} \label{hardsec}
In order for HeapSafe to guarantee protection, it needs to ensure that the HeapSafe hardware is not compromised. Hence, we need to ensure the integrity of the RoCC HeapSafe engine, so that no malicious code can overwrite metadata entries in the metadata table. This is guaranteed to some extent by appropriately setting the RocketChip configuration to run RoCC operations in machine (M) mode only, while rest of the code runs in user (S) mode. This prevents any malicious code to run the RoCC instructions while in user mode. The security of the hardware can be further improved by running the HeapSafe library functions in machine mode only. This requires some additional mediation logic in the application code utilizing traps that requires a switch to machine mode from user mode when calling a HeapSafe function, and then exit to user mode after returning from the function. This can mitigate code-reuse attacks that might try to run the HeapSafe library functions maliciously.

\subsection{PMP vs. HeapSafe} \label{pmp}
The RISC-V architecture provides some basic memory protection as part of the ISA. There are 16 Physical Memory Protection (PMP) registers in the base architecture, which can be utilized to perform access control on different memory regions. The PMP registers allow machine (M) mode to specify which memory regions are available during user (U) mode operations. This is an easy and low-cost way to implement memory protection in user mode for simple systems. However, due to the limited number of registers available, it imposes a restriction on the number of regions it can protect. Hence this is not scalable to more complex applications. Furthermore, PMP protected regions need to be contiguous in physical memory, it can lead to memory fragmentation.

In contrast, HeapSafe can be applied in a more granular manner to individual pointers pointing to memory locations. The number of regions to be protected is not restricted by the core architecture, but set by the configurable HeapSafe engine. Thus HeapSafe is more scalable and versatile than PMP.

\subsection{Backwards compatibility} \label{bccompat}
Our HeapSafe implementation is fully backwards compatible with standard unprotected pointers. This allows the source code programmer to
use a mix of protected and unprotected pointers. The programmer can opt to use HeapSafe protected pointers only for security critical heap regions. Although this is less secure, it improves the performance since the program is not being slowed down due to unnecessary validations.

We achieve backwards compatibility by assuming a \textit{tag} value of 0, since pointers in user-level code has the MSB bits as 0. Since this is inherent to the design, we simply exclude 0 as a \textit{tag} for HeapSafe pointers. If we encounter a pointer with its \textit{tag} bits as 0, we treat the pointer as a non-protected pointer and exclude it from validation.

\subsection{HeapSafe system generation} \label{conf}

\begin{figure}[t]
\begin{center}
\includegraphics[trim=0 4.2in 0 0,clip,width=0.49\textwidth]{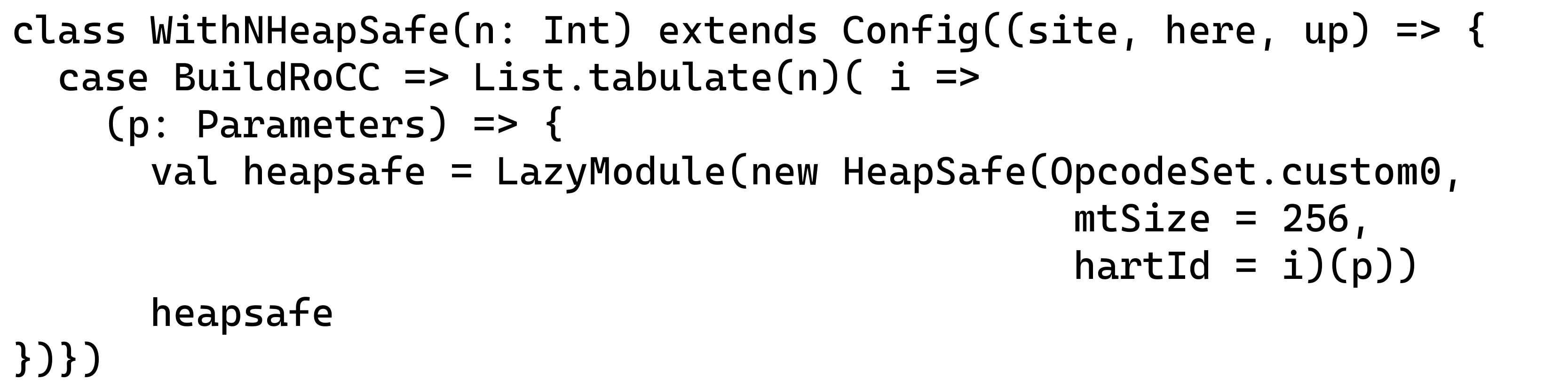}
\end{center}
\caption{RocketChip config class for configurable HeapSafe module generation.}
\label{sysgen}
\end{figure}

Due to the flexibility in HeapSafe's design configuration, we can customize the size of the metadata table to be generated on the hardware. We can also create multiple instances of the HeapSafe coprocessor to support simultaneous multi-process protection. The code  in Fig. \ref{sysgen} demonstrates the easy configurability of HeapSafe design.

When generating a system configuration with HeapSafe, we can set the parameter \texttt{n} to specify the number of HeapSafe modules to instantiate. We associate a \texttt{hartId} to each HeapSafe module. The size of the metadata table can be set through the \texttt{mtSize} parameter. In this config, it is set to 256. In the HeapSafe module implementation, we calculate the bit allocation scheme for the tag in the \textit{safe\_pointer} as $\log_{2} {mtSize}$, e.g., if \texttt{mtSize} = 256, we set the tag bit allocation to the MSB 8 bits in the \textit{safe\_pointer}. 

During the SoC generation, the HeapSafe configuration changes are independent from the core configurations and the rest of the SoC configuration. This provides an advantage that the HeapSafe module can be hooked up to any RISC-V core configuration implementing the RoCCIO interface. As long as the application running on the core is compiled including the HeapSafe library, protection can be provided by the HeapSafe hardware.

\section{Conclusion}
We presented HeapSafe, a customizable and lightweight heap protection hardware engine for the RISC-V ISA. We ensured heap integrity using tagged pointers and enforced metadata propagation for common pointer operations. HeapSafe improved performance over traditional software approaches. The design allows easy configuration and scalability of the security architecture implementation.

\bibliographystyle{ieeetr}
\bibliography{bibliography}

\end{document}